\relax
\documentclass[letterpaper]{article} % DO NOT CHANGE THIS
\usepackage{aaai20}  % DO NOT CHANGE THIS
\usepackage{times}  % DO NOT CHANGE THIS
\usepackage{helvet} % DO NOT CHANGE THIS
\usepackage{courier}  % DO NOT CHANGE THIS
\usepackage[hyphens]{url}  % DO NOT CHANGE THIS
\usepackage{graphicx} % DO NOT CHANGE THIS
\usepackage{graphicx}  % DO NOT CHANGE THIS
\usepackage{subcaption}
\usepackage{hyperref}
\usepackage{amsmath}
\title{Moving with the Times: Investigating the Alt-Right Network Gab with Temporal Interaction Graphs}
\author{Naomi A. Arnold,\textsuperscript{1}\thanks{Corresponding author: \href{mailto:n.a.arnold@qmul.ac.uk}{n.a.arnold@qmul.ac.uk}} Benjamin A. Steer,\textsuperscript{1} Imane Hafnaoui,\textsuperscript{1}  Hugo A. Parada G.,\textsuperscript{2} \\ \Large \textbf{Raul J. Mondrag\'on,\textsuperscript{1} Felix Cuadrado\textsuperscript{2}  and Richard G. Clegg\textsuperscript{1}} \\ 
\textsuperscript{1} School of Electronic Engineering and Computer Science, Queen Mary University of London \\
\textsuperscript{2} Universidad Polit\'ecnica de Madrid \\
}

\begin{document}

\maketitle

\begin{abstract}
Gab is an online social network often associated with the alt-right political movement and users barred from other networks. It presents an interesting opportunity for research because near-complete data is available from day one of the network's creation. In this paper, we investigate the evolution of the user interaction graph, that is the graph where a link represents a user interacting with another user at a given time. We view this graph both at different times and at different timescales. The latter is achieved by using sliding windows on the graph which gives a novel perspective on social network data. The Gab network is relatively slowly growing over the period of months but subject to large bursts of arrivals over hours and days. We identify plausible events that are of interest to the Gab community associated with the most obvious such bursts. The network is characterised by interactions between `strangers' rather than by reinforcing links between `friends'. Gab usage follows the diurnal cycle of the predominantly US and Europe based users. At off-peak hours the Gab interaction network fragments into sub-networks with absolutely no interaction between them. A small group of users are highly influential across larger timescales, but a substantial number of users gain influence for short periods of time. Temporal analysis at different timescales gives new insights above and beyond what could be found on static graphs. 
\end{abstract}

\section{Introduction}
Modern online social networks (OSN) foster different social bubbles, with some platforms hosting opinions that would not be acceptable on mainstream venues. Gab is one such platform, with prominent usage from alt-right supporters. Previous studies found that extremism and radical opinions were a cornerstone of the platform~\cite{zannettou2018gab,lima2018inside}. There are concerns that such online extremism has real world consequences~\cite{johnson2018social,mcilroy2019welcome}. To explore these communities, researchers frequently model a graph from the explicit relationships declared by users, namely followers and group memberships. Since follower relationships can't usually be given an exact starting time~\cite{meeder2011we}, this approach usually involves analysing a static graph which conceals any temporal evolution, e.g. the changing intensity of these relationships over time.

%Our approach, what is the key difference and why it matters
In this paper, we focus instead on the communication happening between users; We study the live interactions captured by the replies of users to posts and messages. This crucial information about individual and collective user behaviour is naturally embedded in the temporal dimension and cannot be observed in static graphs. A key advantage of studying the interactions is that we can observe the active, instant usage of the platform. For such, we model Gab as a temporal interaction graph recording every interaction taking place within the platform and limit the study to that of the structure of the graph formed by the messages, dismissing the content of posts. 

An important property of live interactions is that they might be relevant only during a limited time span across the network's existence. We explore this feature by windowing the Gab temporal interaction graph over multiple temporal resolutions (ranging from one hour to one year at a time). Looking at the graph at multiple temporal resolutions allows us to observe different types of dynamics around user interactions~\cite{saramaki2015seconds}. 

%Examples illustrating why it matters on what we can do
Our approach provides a new way of observing OSN communities. Some researchers characterise OSN as either social (friends interacting) or broadcast (influencers writing to strangers), and the latter may be true of Gab~\cite{zhou2019elites}. We investigate this more deeply by looking at the proportion of interactions between user pairs who had not previously communicated. We also investigate influence on Gab by examining whether a dominant group of most important users (measured by number of interactions) persists or whether this influence is short lived. 

%Wrapup before Dryish contributions, could be cut 
The complete nature of the Gab data makes this the first paper, to our knowledge, that analyses the interaction graph of a complete social network. We believe that viewing temporal networks at a number of timescales is a tool that can provide considerable insight into Gab and other social networks. Consistently, we found that the insights gained from viewing the interaction graph at one timescale were enhanced by the insights from viewing it at other timescales. Hence, we hope that the techniques used in this paper will prove valuable for other OSN researchers. 

%Contributions
This paper makes the following contributions; 1) We present a methodology to analyse OSNs through the lenses of different temporal windows applied to an interaction graph; 2) We explore the Gab interaction network observing trends in its growth over the first eighteen months. We show that the slow growth over long time windows decomposes to rapid rises and falls at the daily and hour timescale driven mainly by events of interest to the user base; 3) We look at the cohesiveness of the Gab userbase, asking whether it forms a single connected group. We find that at a timescale of an hour the network oscillates between being mostly a single connected component at peak hours and being largely fragmented at off-peak hours. We believe this behaviour has never previously been observed; 4) we explore the relative influence of the top Gab users over different time resolutions, observing that a small number of users gather an extremely high level of attention -- several users interact with more than 5\% of the entire user base every month.

\section{Background and related work}

\subsection{Gab Analysis}
Gab~\footnote{http://gab.ai} is a fairly new social media platform, in a lot of ways similar to Twitter, that claims to champion ``free speech, individual liberty, and the free flow of information online". Efforts to open accessibility to datasets with different characteristics of this platform are continuously provided for the research community~\cite{zannettou2018gab,fair2019shouting}. The first venture into studying this network was performed by~\cite{zannettou2018gab}. They found that Gab attracted predominantly alt-right views and conspiracy theorists, in which hate speech and discriminatory language were highly present. 
\cite{lima2018inside} delve deeper into the characterisation of the network by looking into the nature of Gab users and the type of content being shared. In doing so, they link the unmoderated nature of posts to the emergence of echo chambers around alt-right-leaning content.
\cite{kalmar2018twitter} compared the spreading of the Soros Myth on Twitter and Gab and concluded that, in addition to anti-Semitic content being more prominent on Gab, its users display unabashed willingness to post and share such content.
The work of~\cite{mcilroy2019welcome} was motivated by the Pittsburgh synagogue shooting in which they analysed the use of language and the topics discussed and their evolution since the creation of the network until the violent attack. They note a shift from neutral terms and site-setup topics to an increase in more extreme offensive language and racist subjects. They also observe a period in time where the popular topics include non-English terms, an indication to the network expanding to a more European presence.
A recent analysis~\cite{zhou2019elites} further explored this finding by comparing the behaviour of English and German speaking users. They observe that, although small in proportion, the German accounts tend to build tighter connections amongst themselves. A comparison with a Twitter dataset highlighted a more elitist structure to the Gab network with mostly homogeneous content. 
Most of the research presented so far focused on investigating the content from a language perspective, and little has been done to analyse the structure of the network; more interestingly, the type of interactions that govern this particular platform. 
One study~\cite{bagavathi2019examining} examined the nature of conversations on Gab by analysing cascades in the corresponding graph. They notice that cascade conversations tend to emerge from linear interactions and evolve towards more complex structures as topics become viral.
A recent work by~\cite{mathew2019hate} explores the temporal behaviour of the userbase on Gab. By examining the nature of speech, they categorise the users into hateful and non-hateful. They report a fast increase in the number of hateful users and that new users seem to embrace hateful speech quicker. Looking at the structure of the follower graph, they notice that a large number of hateful users are at the core of the network and seem to attract a lot of attention which was steering the conversations towards hateful speech at a faster rate.
In here, we aim at exploring the temporal aspect of the Gab network by focusing on the interactions among the users without looking at the content or the follower relationships but rather at the structure of the interaction graphs.

\subsection{Temporal graphs and windowing}
Graphs are a popular means for modelling relationships between members of online social networks, and are useful for understanding collective behaviours and the spread of information and content on these platforms. There are different models for constructing such a graph depending on the purpose of the study and available data. Much early work in this area focused on static graphs, with edges representing a binary friendship or directed `following' relationship~\cite{mislove2007measurement}. Later work coined the phrase `interaction graph'~\cite{wilson2012beyond}, focusing rather on actual communications between users, with the idea that these might bring more insight into the relationship between users than previously used friendship graphs. Contact graphs are another type where the interaction is associated with a duration~\cite{nicosia2013graph}. A similar concept of an `activity network' was explored in a temporal setting, examining the persistence of interactions between users on Facebook~\cite{viswanath2009evolution} finding that only a minority of declared friendships were maintained with a monthly `wall post' interaction. 

An important question when looking at temporal graphs is that of the appropriate time scale to study. Analysing the evolution of the aggregate graph (every link that occurs before some time $t$) has an important problem: the aggregate graph grows continually and even relatively large events would be eventually eclipsed. An obvious approach is to study the graph of contacts within a particular `window'.
A small number of authors have investigated how the study of a network depends critically on the time scale at which you study it. In~\cite{sekara2016fundamental,stopczynski2014measuring} the authors look at bluetooth contact data for 1,000 individuals. They show that studying the data in five minute windows produced identifiable social groups but aggregating this to a day long network blurred these groups into a single connected component. The authors of~\cite{clauset2012persistence} look at a small contact network (67 individuals) and similarly find considerable variation in network measures using time scales, varying from five minutes to a full day. In~\cite{perra2012activity} the authors consider three graph datasets -- two citation networks and one social network -- and look at the frequency of pairwise interactions over a variety of window lengths. They observed that these datasets make up sparse, disconnected networks for shorter window lengths, with giant connected components forming as the window lengths became longer. Unfortunately, they were only able to do a limited amount of exploration on three different window lengths. In ~\cite{krings2012effects} the authors investigate mobile telephone network data within various window sizes, from one hour to several days, with a particular interest in diurnal behaviour. They used the same null model of shuffled data as this paper and showed that the large connected component could be much larger than in the shuffled model for window sizes of one to two hours. Choice of window size to analyse can affect how systems are built. In~\cite{xie2016exploiting} the authors build delay tolerant networks modelled using contact graph datasets and note that ``too large a window size cannot capture the time-varying features [..] too small a window size may [..] lead to inaccurate calculation of social metrics." This emphasises the need to explore temporal networks at different window sizes since diverse patterns of behaviour would be distinct in one window and obscured in others.
The work by~\cite{leo2015non} offers an upper-limit on the range of possible periods that could be analysed. They looked at whether graph snapshots of a given window size were preserving the propagation properties of the original temporal network. They defined an automatic way to determine a saturation scale by tracking the window that brings the occupancy rate distribution closest to a uniform density distribution. They show that this describes a threshold on the window size where the aggregated graph would lose its temporal dynamics as it relates to properties of propagation if the window was larger than this scale.
The work presented here doesn't aim at determining a single window to perform the analysis but rather explores different window sizes to uncover dynamic network patterns that have different rate of change and necessitate looking at the network changes at multiple scales. 

\section{Dataset collection and analysis}

\begin{table}[tpb]
\centering
    \caption{Details on the userbase and interactions within Gab that span the period between August 2016 -- May 2018.}
    \begin{tabular}{p{0.6\linewidth} p{0.25\linewidth}}
       \textbf{Property}& \textbf{Quantity}\\
       \hline
        Number of users & 169,745\\
        \% of users that never reply & 48.5 \%\\
        Number of posts & 19,091,476 \\
        \% as original posts & 62.5 \%\\
        \% as replies & 37.5 \%\\
        \% as self-replies & 1.53 \% \\
        \hline
    \end{tabular}
    \label{tab:datasum}
\end{table}

This paper investigates the structural characteristics of the Gab social network as they evolve through time by focusing on user interactions. Our starting dataset is obtained from crawling the Gab network by relying on the Rest API provided and consists of 95 GB of raw data. It contains every message and reply posted between 10th August 2016 (the start of the Gab network) and 5th May 2018 (when an API update made it difficult to identify new posts). It omits only those messages that were deleted either by the users who posted them or by moderators. The relevant details of the dataset are summarised in Table~\ref{tab:datasum}. Similar to Twitter, the platform allows users to follow and be followed by other users. However, the time of the following isn't accessible which only allows for static graph analysis. Instead of looking at follower/friendship relationships, we rather explore the interactions between the users in the form of posts. 
Users of Gab share posts, which may consist of text, images or urls, much like Twitter. They can interact by making a  \textit{post} or interact with other posts by \emph{reply post} or \textit{repost} which are henceforth both identified as \textit{replies}. We formally define an interaction as a tuple $(u, v, T)$ where $u$ is a user replying at time $T$ to a post authored by $v$.

\section{Methodology}

\subsection{Windowing and graph construction}
\label{sec:windowing}
\begin{figure}
    \centering
    \includegraphics[width=0.95\linewidth]{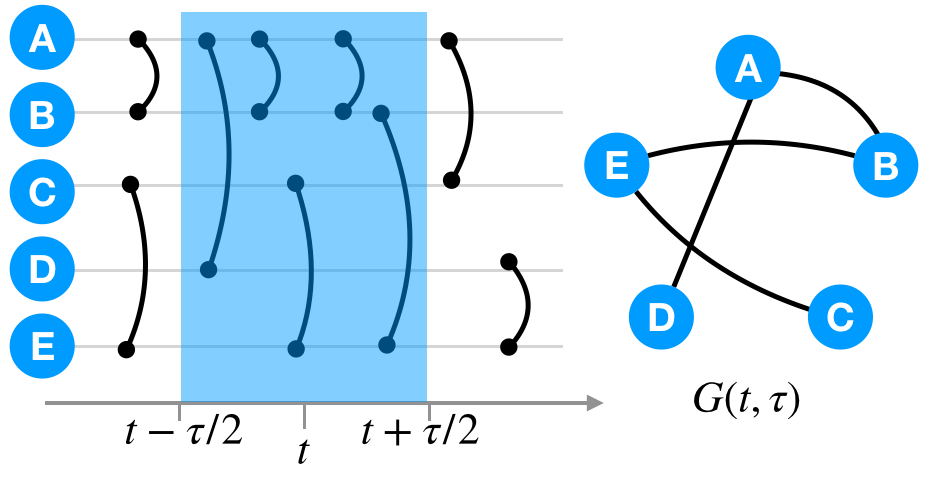}
    \caption{Windowing procedure to construct the graph $G(t, \tau)$.}
    \label{fig:windowing}
\end{figure}

With temporal graphs, snapshots of graphs are generally built at time $t$ which includes every activity that occurred at time $t$. When windowing is concerned, graph snapshots within a window of length $\tau$ are aggregated together to form a single graph. 
Formally, a graph $G(t,\tau)$ at time $t$ and window size of $\tau$ is constructed by adding a single directed edge $e_{uv}$ between users $u$ and $v$ if an interaction $(u,v,T)$ occurs for $T \in ( t - \frac{\tau}{2}, t + \frac{\tau}{2}]$. This means necessarily that for $0 < \tau_1 < \tau_2$, the graph $G(t,\tau_1)$ is a subgraph of $G(t,\tau_2)$. Figure~\ref{fig:windowing} illustrates an example of such windowed aggregation. Notice that, although users A and B interact more than once during this interval, only one edge is added between them. Also, only users participating in an interaction during a window are included as a node in the graph. To observe temporal patterns throughout the history of the network, we construct graphs by sliding windows of size $\tau$ and an offset $\Delta$. Unless stated otherwise, for the rest of the work, we use the window sizes of $\tau \in \{1 \text{ hour}, 1 \text{ day}, 7 \text{ days}, 30 \text{ days}\}$ and offset $\Delta = 1 \text{ day}$.

To enable such high-granularity analysis, we employ an efficient graph processing system called Raphtory~\cite{steer2020raphtory}. Raphtory models ingested data as a temporal property graph, where the full structural and property history of entities within a network are maintained in-memory. This allows the user to view the exact state of a graph at any point throughout its evolution. A user may additionally apply a `windowed view' which only returns entities that have been updated within the window size, looking back from the chosen view time. In this fashion, the scraped Gab data was ingested and built into the interaction graph described above. Windowed views were then produced at each day throughout its lifetime, applying the chosen window sizes. For several of the results below, this translates into over 17,000 windowed views of the graph being materialised. This is handled automatically by the tool making it possible to run different analysis algorithms on varied temporal scales and observe patterns that may be hidden with larger window sizes.

\subsection{Randomised reference models}\label{sec:rrm}

We augment our analysis where appropriate with the use of two randomised reference models to investigate potential structural and temporal factors determining the phenomena we observe. Our first reference model is a ``shuffled timestamps" model~\cite{gauvin2018randomized,holme2012temporal}. This model randomly shuffles the timestamps $T$ of all interactions $(u,v,T)$ while keeping $(u,v)$ fixed. In this way, the graph obtained by aggregating all interactions within the dataset is conserved within this null model, as well as the set of interaction times, but removed are the temporal correlations between interactions from adjacent pairs of users. These correlations are common to interaction networks, arising, for example, as burst trains~\cite{karsai2012correlated} (sequences of repeated interactions between node pairs) or as temporal subgraph motifs~\cite{zhao2010communication}. The second reference model we use, commonly known as the restrained randomisation model~\cite{maslov2002specificity}, generates a random graph snapshot with a given degree sequence, usually chosen to match that of the dataset. After creating a graph for a specific window we use this model to create a graph with the same degree sequence. This may be used to test whether observed graph features can be explained just by the degree sequence of the network or whether structural correlations such as assortativity or community structure play a role.

\section{Results}

\begin{figure}[t!]
    \centering
    \includegraphics[width=0.95\linewidth]{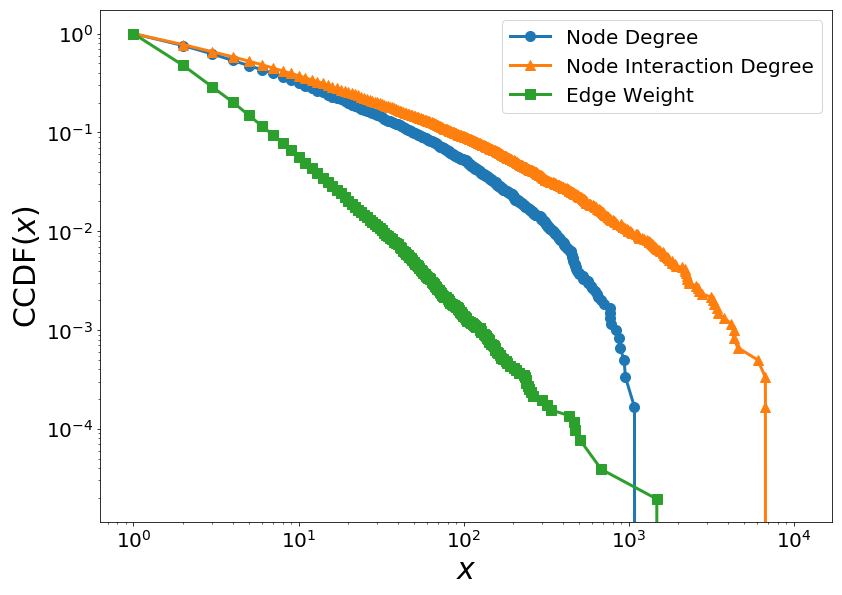}
    \caption{Distributions of node degree (the number of unique users a user interacts with), node interaction degree (the number of interactions a user is involved with) and edge weight (the number of times a given pair of users interacts).}
    \label{fig:degdist}
\end{figure}

\begin{figure}[b!]
    \centering
    \includegraphics[width=0.95\linewidth]{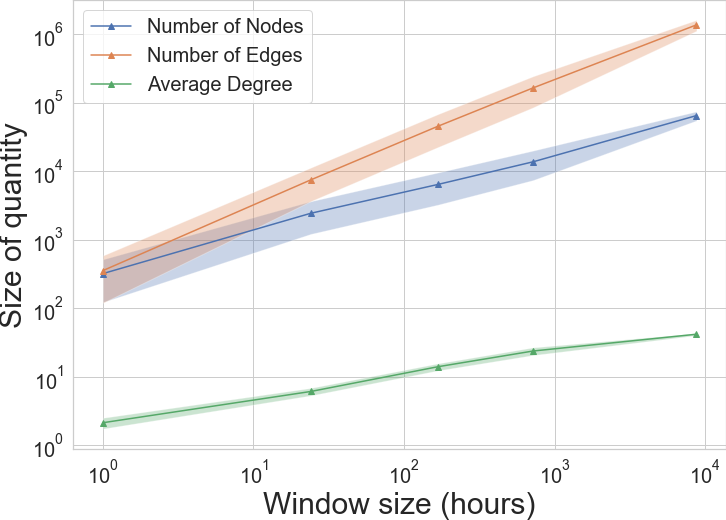}
    \caption{Average value of the number of nodes, edges and average degree as the window size grows; the shaded region represents one standard deviation above and below the mean.}
    \label{fig:windowscale}
\end{figure}

\begin{figure*}
    \centering
\begin{subfigure}{.49\textwidth}
    \centering
    % include first image
    \includegraphics[width=\linewidth]{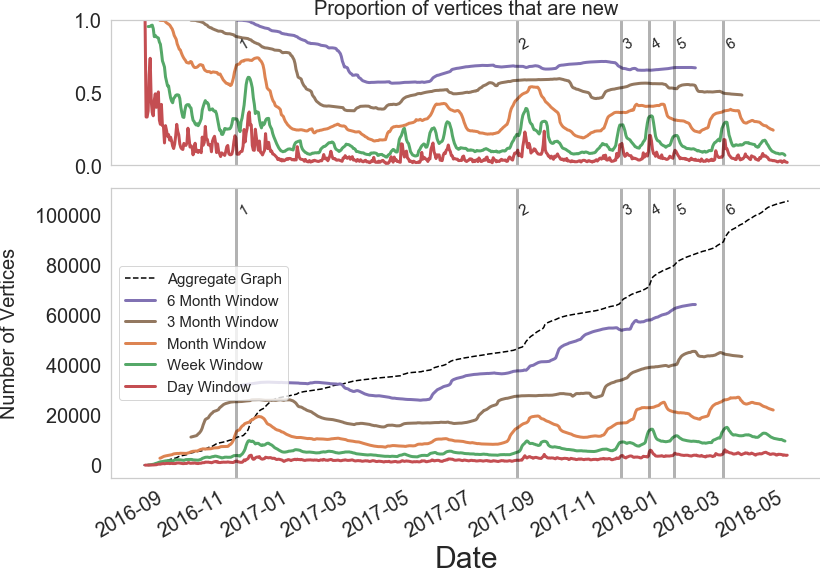}  
    \caption{Number of users}
    \label{fig:no_nodes}
\end{subfigure}
\begin{subfigure}{.49\textwidth}
    \centering
    % include second image
    \includegraphics[width=\linewidth]{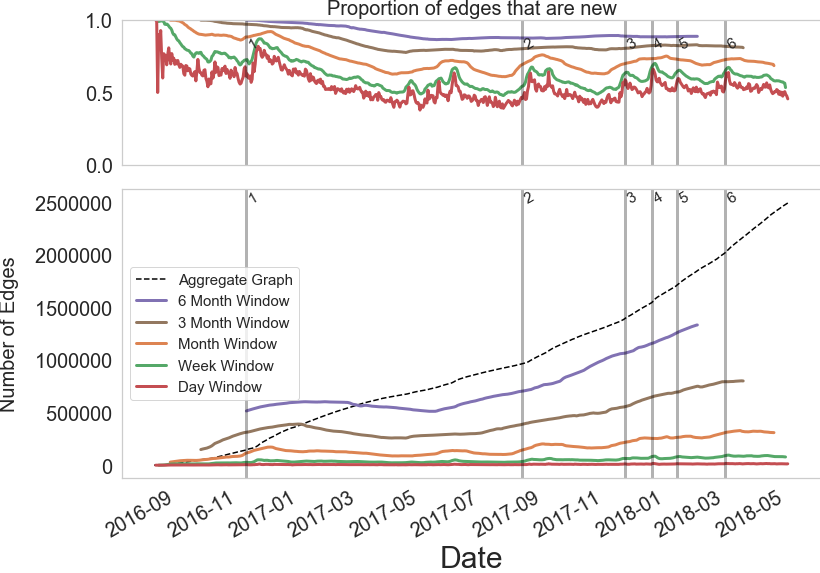}  
    \caption{Number of edges}
    \label{fig:no_edges}
\end{subfigure}
\caption{Number of (a) users and (b) edges between unique user pairs across different window sizes (bottom) and the proportion of these that are new (top). `New' here means that this is the first window in which they are present. Shown also as a reference is the number of nodes and edges in the aggregate graph, i.e. from aggregating the interactions from time 0 to time $t$.}
\label{fig:fig}
\end{figure*}

Our investigation of Gab considers three important questions about the network. \emph{(i) What is the nature of growth in the Gab network and what drives that growth?} We address this question by looking at how the number of nodes and edges in the interaction network changes over time and the proportion of those nodes and edges that have never been seen before.
\emph{(ii) Is the Gab network a single connected group of interacting users or a number of disconnected groups that rarely interact with each other?} We consider this by investigating the behaviour of the large connected component for the graph. Finally \emph{(iii) we ask whether Gab is dominated by a consistent group of influential users or whether the most influential users change over time}. We address this by looking at the most connected users in the network. 

As this paper focuses on interactions rather than declared friendship/follower relations, we first examine how these interactions are distributed across the network, with Figure~\ref{fig:degdist} showing some of these quantities pertaining to interactions. Node degree refers to the number of distinct users that node has interacted with, whereas interaction degree refers to the total number of interactions that node has had. Edge weight refers to the number of interactions between each pair of users, excluding pairs which never interact. We notice that 52\% of the user pairs that interact do so once and never again, with a long tail of few edges over which many interactions occur. 

In all of these investigations, we analyse how the network evolves as time goes on but also how the network quantities change as we consider different sizes of time window. The latter parameter is important; the larger the time window to investigate the network, the more interactions we will be looking at. Figure~\ref{fig:windowscale} shows the change of some basic network properties when varying the size of the window. As would be expected, larger windows include a greater number of nodes and edges. A less intuitive but well-known observation for growing networks is that the longer the time period considered, the higher the average degree (or edge/node ratio); a process sometimes called densification~\cite{leskovec2005graphs}. A similar pattern was observed when using window based analysis to study proximity networks~\cite{clauset2012persistence}.

%Analysis flow: We study interaction graph dynamics at different temporal resolutions
%\begin{enumerate}
%    \item First view at the different windows, can we give some general global info? Number of vertices/edges is probably the most straightforward
%    \item Looking at inter-window changes, in both vertices and edges. In general, then zooming in on particular events that we can differentiate from the `norm'
%    \item Now looking at individuals from each window (top 20 in in/out degree rankings). What can we say from them? what does Jaccard tell?  
%    \item moving to topology, how are the interactions taking place? connected components. These reflect in an interaction network how closely do users interact on discussing the same topics
%\end{enumerate}

\subsection{Drivers of Growth in Interactions}

An aggregate graph viewed over time can only ever add nodes and edges and may not reveal clear patterns in the temporal behaviour of these entities. In this section, we use different window sizes to explore whether we can identify factors affecting growth or decline in these interactions. We use the definition of nodes and edges within a window $\tau$ which make up the graph $G(t,\tau)$ defined in the previous section. We distinguish a ``new'' node or edge within a window as entities that have never been seen at any time before this window. In other words, these represent users and pairwise interactions that occur in the window  $T \in ( t - \frac{\tau}{2}, t + \frac{\tau}{2}]$ and were never active before $( t - \frac{\tau}{2})$. Adjusting the window size $\tau$ provides  different definitions of who is an ``active'' user. We investigate whether the system is driven by returning users who continue to contribute over time or new users who have never been seen before the time window under analysis.

Figure~\ref{fig:no_nodes} shows the number of users according to daily, weekly, monthly, quarterly and 6-monthly window sizes (bottom), and the proportion of those users that are new (top). In addition, the running total (aggregate) number of users is plotted as a reference. The number of users that are active monthly or more regularly makes up a small proportion of the total userbase, which is not surprising given the prevalence of bots and duplicate accounts on social networks~\cite{silva2013botnets}. On average 19\% of monthly users are active on a given day, compared to 44\% for Twitter~\cite{twitterreport2019}.

Similarly, Figure~\ref{fig:no_edges} shows the number of edges and the proportion of new  interactions in different windows. Surprisingly, the majority of communications in most windows of a day or greater are between users who have never previously interacted. The ratio holds even in the later months of our dataset where the mass of users was higher. This is in contrast to work~\cite{grindrod2011social} which has found that many edges in temporal social networks tend to be persistent, with the probability of an edge returning depending positively on whether it exists already. They observe that 12\% of weekly edges are new interactions, a figure which stays low throughout the dataset (after the first week). This is in contrast to an approximate proportion of 60\% we see in the Gab data. It does, however, add weight to studies~\cite{zhou2019elites} that have characterised interactions on Gab as broadcast behaviour rather than engaging in conversations; where interactions exist not to build up social relationships but to reinforce hierarchies.

\begin{table}[tpb!]
     \centering
     \caption{Events coinciding with an influx of new users and interactions on Gab.}
     \begin{tabular}{c l p{0.67\linewidth}}
        \textbf{\#} & \textbf{Date} & \textbf{Event} \\
        \hline
        1 & 09/11/16 & Trump wins election \\
        2 & 11/08/17 & Unite the Right rally \\
        3 & 21/11/17 & Plan to repeal neutrality announced \\
        4 & 18/12/17 & Twitter suspends white nationalists \\
        5 & 12/01/18 & Trump `sh*thole countries' comment \\
        6 & 01/03/18 & Gilmore lawsuit against Alex Jones \\
        \hline
    \end{tabular}
   \label{tab:changepoints}
\end{table}

Multiple window sizes help in identifying events that are either hidden with larger window scales or lost within the noise of smaller windows. With this particular dataset, the weekly ratio seems to provide the most clear indicator of bursty spikes of new users. We have compared these peaks with key events from the alt-right movement. A list of some events that could be triggers for these activity bursts have been marked in Figures~\ref{fig:no_nodes} and~\ref{fig:no_edges} with a key in Table~\ref{tab:changepoints}. We consolidate these beliefs by measuring the frequency of posts containing a given keyword on the event day, compared with a corpus of 250,000 randomly selected posts. As a benchmark comparison, `Trump' is mentioned 3.5 times more on the day after the 2016 US election than on a randomly chosen day.

Some growth bursts seem to centre on events around the Trump administration, with the site's largest peak in proportion of new users after the site's birth coming shortly after Trump's election (event 1). Trump's remark referring to Haiti, El Salvador and African nations as `sh*thole' countries (event 5) coincides with one of these. We found the word `sh*thole' appearing 766 times more on the selected day than in the reference data; pushing it to be found in~\cite{mcilroy2019welcome} as one of the all-time top 15 over-represented words on the platform. Some peaks in new users seem to be driven by migration from Twitter in response to clampdowns on the alt-right, the site founded shortly after the high profile banning of Milo Yiannopoulos from Twitter in July 2016. One such clampdown is on 18 Dec 2017 (event 4) when Twitter removed numerous accounts associated with the far right, (including for example, Paul Golding and Jayda Fransen of Britain First who joined Gab the same day) prompting a 2.2 times increase in occurrences of `Twitter' among Gab posts. The announcement to repeal net neutrality (event 3) sees a spike in edges, 'neutrality' yielding a 2.2 times relative frequency increase. What seems to be the site's largest and most sustained increase in number of users after the birth of the site is around the Charlottesville Unite the Right rally (event 2) and its aftermath. The number of active users in this month roughly doubles from a month previous, with over half of these users never seen before on the site at its peak. In March 2018 (event 6) a lawsuit is filed against Infowars' Alex Jones, one of the top 20 most followed users of Gab~\cite{zannettou2018gab} regarding a Charlottesville conspiracy he perpetrated, prompting another spike of discussions on the site. 

\subsection{Considering the Gab Userbase as a Connected Community}

\begin{figure}[tbp!]
    \centering
    \includegraphics[width=0.95\linewidth]{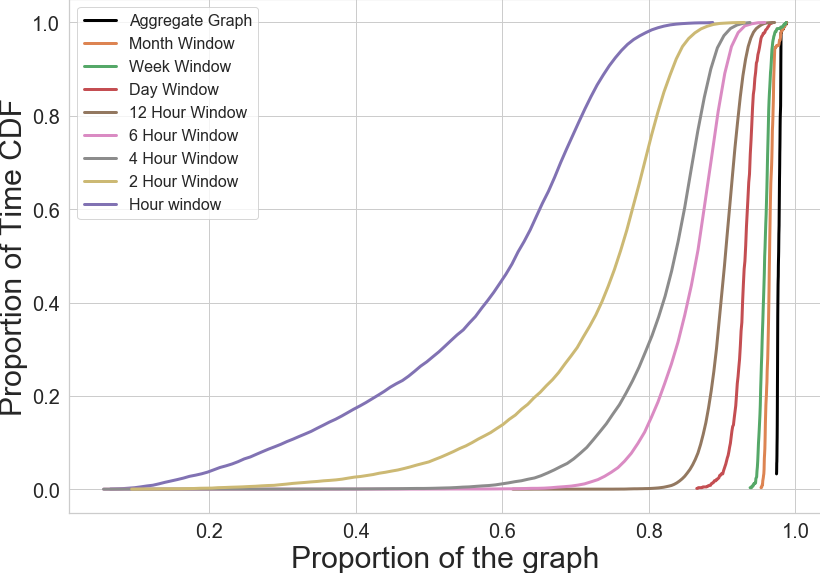}
    \caption{CDF of the size of the largest connected component (LCC) as a proportion of the total graph for each window size. A point on this graph shows the proportion of the time (y-axis) the proportion of graph within the largest connected component was smaller than the value on the x-axis.}
    \label{fig:cdf_lcc}
\end{figure}

The Gab userbase is characterised by homogeneity, especially driven by political topics. It is useful thence to investigate the extent to which they form a single community\footnote{Here we refer to a community in the informal sense and not to the mathematics of community detection.}. We might consider the ties among the Gab community by looking again at the data in Figure~\ref{fig:no_edges} and in particular the proportion of new edges. The day trace for new edges shows that 40\% to 70\% of daily interactions are between users who have never interacted before that day. This is consistent with the idea that Gab is not well characterised as a community of friends who consistently interact with each other.

We address this question by observing the size of the largest connected component (LCC). Aggregate graphs of OSN almost always show that the majority of users in a network are part of a large connected component. Considering time windows of data gives a much more detailed insight. Smaller window sizes imply fewer nodes and fewer links (analogous to edge removal for graph robustness~\cite{albert2000error}).

\begin{figure*}[tb!]
    \centering
    \includegraphics[width=0.95\linewidth]{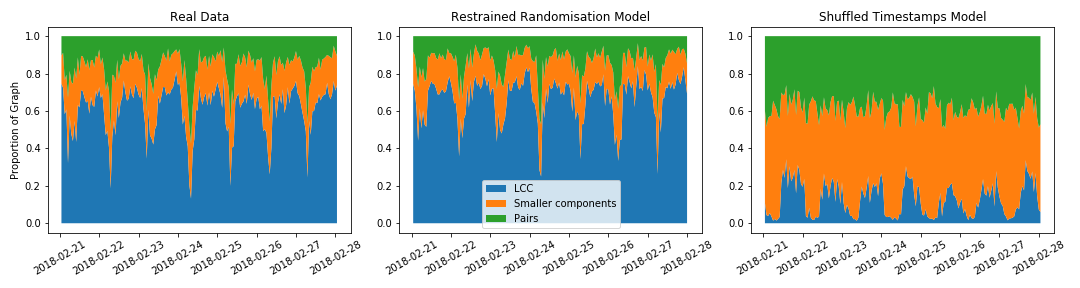}
    \caption{Comparison of the relative size of the largest connected component, components of size 2 (pairs) and intermediate sized components when the graph is viewed with an hour sized window. Values are given for the same week time period, in the real data (left), the restrained randomisation model (centre), and the shuffled timestamps model (right). Graphs are obtained over the hourly window for a week period.}
    \label{fig:proportion}
\end{figure*}

\begin{figure}[tb!]
    \centering
    \includegraphics[width=0.95\linewidth]{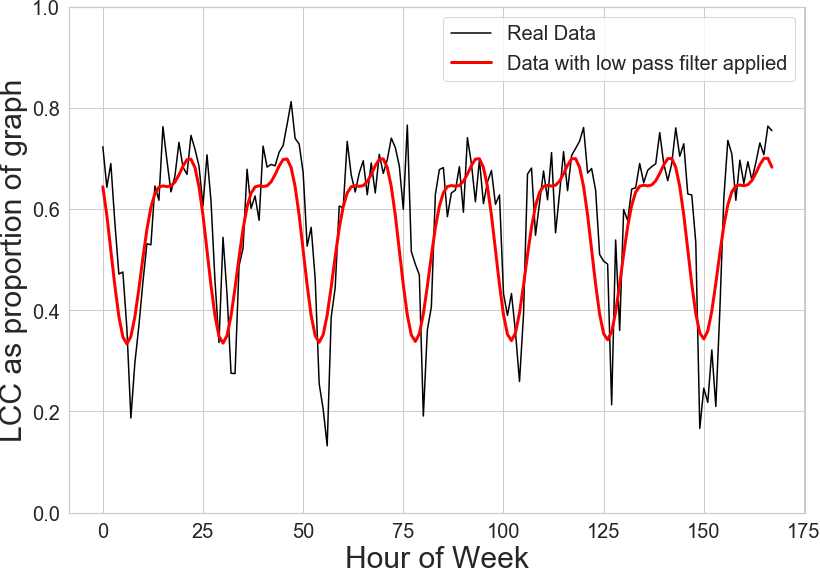}
    \caption{A week of data showing the proportion of the graph that is within the largest connected component when the graph is viewed in hourly windows. The red line smooths this with a low pass filter. With the hourly window most of the graph is in a single connected component at peak hours but at off peak hours the graph ``shatters" into smaller components.}
    \label{fig:timeseries}
\end{figure}

Figure~\ref{fig:cdf_lcc} shows the proportion of users connected by the LCC for different window sizes $\tau$. For example, at a window size of one day (and all longer windows) the LCC connects 90\% of users for the vast majority of windows (the LCC proportion is above ninety percent nearly one hundred percent of the time). However, the smaller the window size we observe, the higher variability we find in the size of the LCC. As might be expected the number of users (and hence the absolute size of the LCC) varies considerably with time of day when studied at the hour level. This is consistent with~\cite{krings2012effects} looking at the same effect in a telecommunications network. What is less predictable is what will happen when we look at the \emph{proportion} of users who are within the LCC.

It is not obvious whether an OSN will form a single LCC when viewed at the finest time scale (one hour). In fact, at the hour time-scale, the Gab network moves between two different regimes. This can be seen in Figure~\ref{fig:timeseries} where we focus further on the proportion of users in the LCC using the hour long window. When we zoom in on a short period of time the reason for the variability becomes clear. The proportion of users who are part of the LCC varies hugely from the peak to off-peak. Gab usage is highly diurnal, driven by a userbase that is largely US- and Europe-based~\cite{zhou2019elites}. At peak hours, 70 -- 80\% of active users are part of a single LCC. At off-peak hours, the LCC contains no more than 30\% of users and Gab becomes several smaller completely disconnected networks that can be thought of as isolated groups (sometimes just user pairs) talking among themselves. We believe that this daily shattering and reforming of the LCC has never previously been observed in OSN data.

 We look a little closer at the nature of activity arising from the diurnal habits of the userbase by applying a low-pass filter to the data in Figure~\ref{fig:timeseries}. The red line represent the data with a low pass filter keeping only the lowest frequency components. We notice twin peaks, a distinct peak with a slightly smaller peak lagging behind. This pattern could arise from two superimposed groups of slightly different sizes that are six hours apart. This is consistent with the work~\cite{zhou2019elites} which found a German and French presence alongside English language content on the platform.

The change of proportion and size of the LCC is not simply an effect of the number of active users at a given time as can be seen by comparing against randomised models. Figure~\ref{fig:proportion} shows the proportion of the graph taken up by the LCC (blue), other components with a size strictly greater than two (orange) and pairs of users only interacting with each other (green). This view shows us not only that the proportion of users in the largest connected component enlarges and shrinks again over the course of a 24 hour period but that at off-peak hours this shrinkage comes mainly with an increased presence of isolated interactions in the network as opposed to more intermediate sized components. First, ignoring any temporal aspect, we test whether the behaviour of the connected component is owing to any structural correlations within the network, using a restrained randomisation model (see section~\ref{sec:rrm}) as a reference (middle panel of Figure~\ref{fig:proportion}). The restrained randomisation model produces very similar results to the real data. However the proportion of users in the LCC is slightly lower and the proportion of users making pairwise interactions slightly higher in the real data, particularly in the off peak. The majority of the behaviour is a result of the number of interactions each present user is engaged with but some arises from Gab's network structure which favours pairwise interactions over group discussions. 
\begin{figure*}[tb!]
    \centering
    \includegraphics[width=0.9\linewidth]{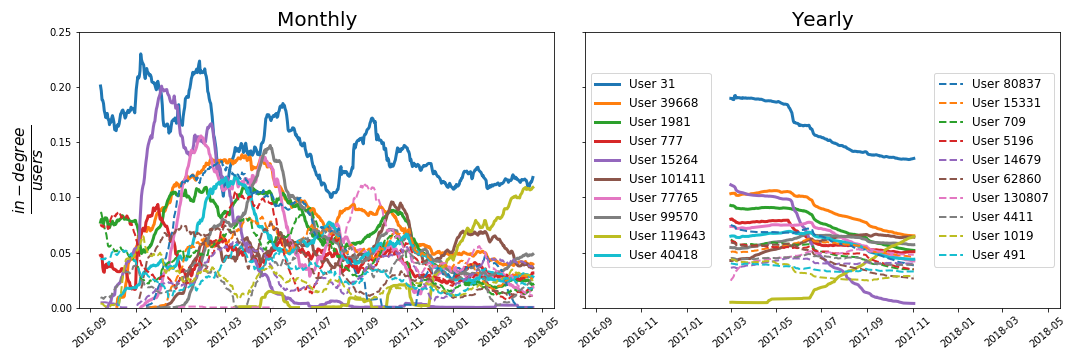}
    \caption{In-degree for a set of top-twenty users for the month (left) and year (right) window sizes normalised by the number of nodes within that window. The top-twenty are chosen by their in-degree in the aggregate graph.}
    \label{fig:degcentrality}
\end{figure*}

\begin{figure}[tbp!]
    \centering
    \includegraphics[width=0.95\linewidth]{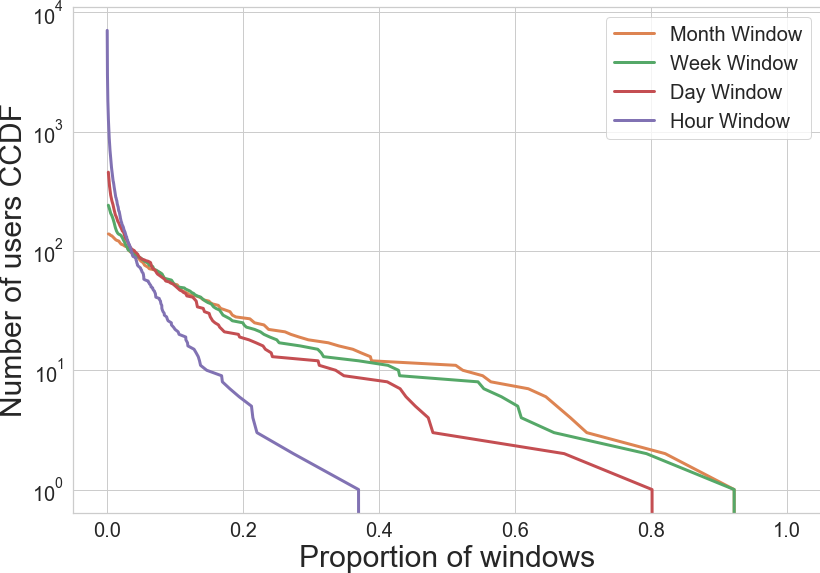}
    \caption{The plot shows for each time window how many users are in the top twenty for a given proportion of the windows of that size.  For example, in the month window 10 users are in the top 20 for 50\% of the month long windows.}
    \label{fig:ccdf_proportion}
\end{figure}

We investigate the role of temporal correlations by considering a randomised timestamps model (see section~\ref{sec:rrm}) which takes the original interaction dataset and permutes the timestamps. In this case Figure~\ref{fig:proportion} (right) we observe that a periodic behaviour persists but that the proportion of users in the LCC is very small for all time periods. The periodicity can be explained by the fact that the higher rate of interactions during peak hours means increased densification and hence a larger LCC. However, the activity rate alone does not explain the proportion of users present in the LCC in the real data during peak hours (the LCC is a much smaller proportion of the graph in the shuffled model), meaning that temporal correlations between interactions are indeed the vital driver in explaining this behaviour.

\subsection{Churn of Influential Users on Gab}

 In this section, we ask how many users \emph{ever} become influential on the platform, and for how long that status lasts. We use the in-degree (number of interations from distinct users) a user receives as a coarse measure of their influence in that window, and we extract the top 20 users for each window. Figure~\ref{fig:ccdf_proportion} reports the proportion of windows a user spends in the top 20 at different timescales. The same nine users that are found in the top 20 in more than half of the month windows are found in nearly half of the week windows, and 30\% of the day windows. When looking through a timescale of a day, we see a large pool of more than 800 different users enter the top 20 at this timescale. For the most part, however, this influence is very short lived, with nearly half of these entering the ranks once and never again.

We also examine trajectories of the top ranked users by in-degree (Figure~\ref{fig:degcentrality}, showing more explicitly the rise (and sometimes fall) of a subset of these users. The users shown are the 20 with the highest in-degree in the aggregate graph. Some of these users hold and maintain a huge level of attention, with the top ranked user~\footnote{Andrew Torba, founder of Gab} in the year window receiving replies from 20\% of the active users within that year and the rest of the users presented in the year window each achieving attention from over 4\% of the network. The year window shows on the most part a stable hierarchy of top users who only occasionally overtake each other, with a downward trend which is explained by the growing nature of the network. The monthly plot shows in more detail when and in what manner each user rises to prominence. For example, user (31) who tops the year window is dominant for nearly the whole time period for the smaller window as well. On the other hand, some users who obtain prominence in the year window achieve this through a short burst of fame seen only through the lens of smaller timescales (seen with Users (15264) and (130807)). Not included are the plots for the day and week windows size which are mostly too noisy to draw out any salient user trajectories. We did note, however, a small number of users who gained a high amount of influence for just a single day each and never again.

\subsection{Summary of main findings}

This paper analysed the first eighteen months of Gab data focusing on the structure of the temporal interaction network. Looking at data through the multiple views allowed by different window scales provides insight that could not be gained from analysing a graph using just one time scale. Even a basic question like ``is the Gab network growing or shrinking?" depends crucially on the time window we look at. When considered in small time windows of hours or days, the userbase for Gab is driven by spikes of arrivals that may be associated with events of importance for potential users of this network. We tentatively identify some real-world events that may be associated with growth in this network, for example, the suspension of some far-right accounts on twitter and the `Unite the Right' rally in Charlottesville. Over longer periods of time the network seems to be growing its number of users albeit extremely slowly.

Considering the question as to whether the Gab userbase is a unified `community' that actively interacts with each other, we show that at peak times the community viewed over a single hour does form a large connected component but in the off peak, with less usage, it becomes a number of much smaller connected components that are completely isolated from each other.

The Gab userbase is highly dynamic. In a given month, typically more than one quarter of the users interacting with the network have never previously done so. At every time scale we consider, 40\% of interactions are between pairs of users who have never previously interacted. The most interacted with users in Gab form a small core of users who dominate most of the time receiving extremely high levels of attention, and much wider pool who gain influence fleetingly. Typically, like the saying `fifteen minutes of fame', users in this wider pool may enter the top twenty for a single time period and then rarely, if ever, reappear. The network appears to be characterised by churn, both in new users appearing and in different users dominating the interactions.

\section{Conclusions}

This paper is the first we know about to analyse a temporal interaction graph for a complete social network dataset (in this case, Gab). We investigate the temporal aspect using a variety of timescales (windows). One might ask whether it is possible to identify an ideal timescale for the study of networks; we have found instead that interesting insights can be gained by looking at a multitude of timescales. By viewing how user-to-user interactions evolve at different timescales we gain unique insights into the nature of the network. While Gab grows relatively slowly over longer time scales, this growth is also characterised by sudden rises and falls over short timescales associated with external events of interest to the alt-right community. We have shown how, at every timescale, Gab interactions commonly (more than 40 percent of the time) take place between `strangers', that is users who never previously interacted.  When considering the Gab system as a community, for longer timescales the majority of the active userbase forms a single connected component. However, analysed at short timescales like an hour this largest connected component dwindles to be a very small proportion of the network in off-peak hours. When looking at the most influential users we see a small elite of users who command an unusually large degree of attention from the userbase. Several users interact with more than five percent of the userbase in any given monthly window. 
In future work we plan to extend our approach to other OSNs where the full set of interactions is available (unlike the sampled data available for Twitter). We will also assess the discrimination potential of different timescales depending on the dataset characteristics.

\bibliography{windowingbib.bib}
\bibliographystyle{aaai}

\end{document}